\documentclass[global,twocolumn]{svjour}

\usepackage{graphicx}% Include figure files

\usepackage{bm}% bold math
\usepackage{amsmath}
\usepackage{amsfonts}
\usepackage{amssymb}
\usepackage{color}

\newcommand{\br}[1]{\langle #1|}
\newcommand{\ke}[1]{|#1\rangle}

\newcommand{\da}{^\dagger}

\newcommand{\vect}[1]{\vec{#1}}

\newcommand{\fax}[1]{{fax: \tt#1}}

\begin{document}

\title{Cavity cooling of translational and ro-vibrational motion of
molecules: \emph{ab initio}-based simulations for OH and NO}
\author{Markus Kowalewski \inst{1} \and Giovanna Morigi \inst{2}
\and Pepijn W. H. Pinkse\inst{3}
\and Regina de Vivie-Riedle\inst{1}}

\institute{
Departement of Chemistry, Ludwig-Maximilian-Universit\"at M\"unchen, Butenandt-Str. 11,
D-81377 M\"unchen, Germany,\\ \fax{+49-89-218077133}, \email{Regina.de\_Vivie@cup.uni-muenchen.de} \and
Departament de Fisica, Universitat Autonoma de Barcelona, E-08193 Bellaterra (Barcelona), Spain \and
Max-Planck-Institut f\"ur Quantenoptik, Hans-Kopfermann-Str. 1, D-85748 Garching, Germany}

\titlerunning{Cavity cooling of translational and ro-vibrational motion of molecules}
\authorrunning{M. Kowalewski et al.}

\date{Submitted 03.08.2007}

\maketitle

\begin{abstract} We present detailed
calculations at the basis of our recent proposal for
simultaneous cooling the rotational, vibrational and external
molecular degrees of freedom ~\cite{PRL07}. In this method, the
molecular rovibronic states are coupled by an intense laser and an
optical cavity via coherent Raman processes enhanced by the strong
coupling with the cavity modes. For a prototype system, OH, we
showed that the translational motion is cooled to few $\mu$K and
the molecule is brought to the internal ground state in about a
second. Here, we investigate numerically the dependence of the
cooling scheme on the molecular polarizability, selecting NO as a
second example. Furthermore, we demonstrate the general
applicability of the proposed cooling scheme to initially
vibrationally and rotationally hot molecular systems.

\end{abstract}

\section{Introduction} Control at the quantum level over both the
internal and external degrees of freedom of gas-phase molecules
has been pursued by several groups worldwide~\cite{quovadis}. This intense
activity has various ultimate goals, such as the realization of
Bose-Einstein condensation of complex systems, the emergence of
new ultracold chemistry \cite{Krems05}, and the applications for high-precision
measurements~\cite{Daussy99,Tarbutt04} and quantum information processing \cite{DeMille02,Tesch02,Andre06}.
The efficient preparation of molecules into predetermined quantum
states is the prerequisite for control. Various
approaches are currently applied. Established methods for
generating ultracold alkali dimers use photoassociation and, or in
combination with, Feshbach resonances~\cite{Feshbach}. Another
approach is based on buffer gas cooling, in which the molecules
are thermalized with a cold buffer gas~\cite{Weinstein98}.
Decelleration of molecules from supersonic nozzles~\cite{BethlemPRL99,Barker},
filtering from an effusive source~\cite{RangwalaPRA03} and collisional
techniques~\cite{Chandler03,Loesch07} are other methods which have been
developed to produce samples of cold molecules.

The application of laser cooling techniques, which had tremendous
impact on atomic and optical physics~\cite{Tannoudji98,Chu98,Phillips98},
to molecules turned out to be inefficient so far.
Here, multiple scattering channels, due to
the molecular ro-vibrational structure, are coupled by spontaneous
emission, leading to uncontrolled heating of the system as an
undesired side effect. Even when heating can be suppressed, the
resulting cooling time is long and may only be feasible for
molecules which are confined in external traps for very long
times~\cite{Drewsen04,Schiller05}. Proposals aimed at
overcoming this problem have been suggested: Sophisticated
laser-cooling schemes, based on optical pumping the
ro-vibrational states~\cite{Stwalley,Drewsen} and optimally
modulated femtosecond excitation
pulses~\cite{Tannor,Q-Control-Molecules}, achieve efficiencies
which are still severely limited by spontaneous decay and its
inherent long time duration.

An alternative way to optically cooling external degrees of freedom of
molecules was proposed and studied in~\cite{Horak,Vuletic00,Lev07,Lu07}, and it makes use
of the enhancement of stimulated photon emission into the cavity mode 
over the spontaneous decay. This mechanism was successfully
applied for cooling the motion of atoms~\cite{Maunz04,Nussmann05,Vuletic03}.
In a recent work~\cite{PRL07}, we extended this idea and proposed 
to optically cool the external as well as the {\it internal}
molecular degrees of freedom by using the coupling to a cavity.
In the proposal the molecules are cooled by a Raman process
in which photons from an intense laser are scattered into
the modes of a resonator by molecular dipole transition,
thereby cooling the motion to the cavity linewidth and the
ro-vibrational degrees of freedom to the ground state.
The dynamics is numerically simulated for the OH radical as a
prototype system and for a realistic set of experimental parameters,
while the molecular properties are
calculated with state-of-the-art {\it \emph{ab initio}} calculations. Starting
with an internal temperature of 300 K we could show that
the translational motion is cooled to a few $\mu$K and the internal
state is cooled to its ro-vibrational ground state~\cite{PRL07}.

In this article we provide the details of the theory and of the
\emph{ab initio} results at the basis of \cite{PRL07}. In order to
demonstrate explicitly the general applicability of the proposed
scheme, we present further numerical simulations where we apply it for the simultaneous cooling of
vibrational and rotational degrees of freedom for OH in
a set up with an artificial high temperature,
such that several vibrational states are initially occupied. Moreover, we
simulate cooling of a similar diatomic system, NO, whose
$\pi$ bonding allows for a larger electronic polarizability. 

This paper is structured as follows. First the central idea is introduced
in section~\ref{idea}, followed by a detailed account of the
theoretical model including the basic equations in section~\ref{model}.
In section~\ref{QChemistry} the two selected molecules OH and NO are
introduced and their properties are evaluated with \emph{ab initio} methods.
In section~\ref{Experimental} the experimental boundary conditions
are given which are the final ingredients leading to the results
presented in section~\ref{Results}. The paper closes with a summary, conclusions and outlook.

\section{Optically cooling molecules to the ground state}
\label{idea}

\begin{figure}[b]
\begin{center}\includegraphics[width=0.45\textwidth]{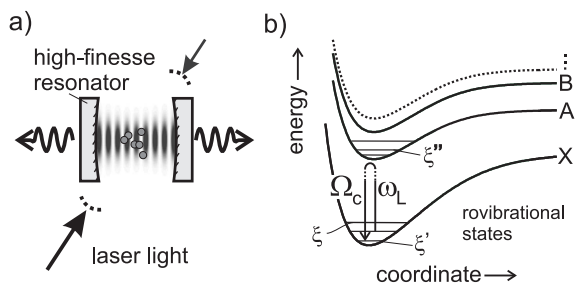}
\end{center} \caption{The idea in a nutshell: (a) molecules which are
assumed to be confined in a high-finesse cavity interact with strong CW
laser light. The laser intensity might be enhanced by a build-up cavity.
(b) The laser light is far detuned from all direct electronic
transitions from the ground state X to electronically excited states
A,B,... in the molecules.The ro-vibrational energy of the molecules
is reduced by a sequence of Raman transition, where laser photons
are scattered into the cavity modes.} \label{Fig:1}
\end{figure}

\begin{figure}[h]
\includegraphics[width=0.45\textwidth]{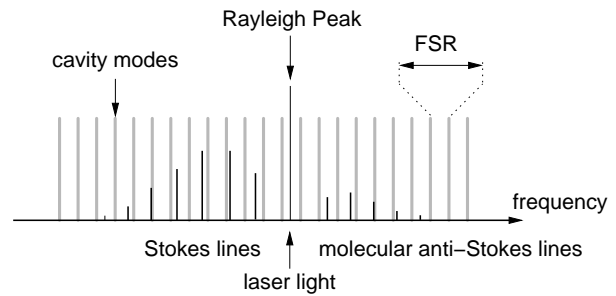}
\caption{Combined with the cavity modes, the
laser drives a sequence of coherent Raman anti-Stokes
transitions which on average reduces
the ro-vibrational energy of the molecules. The spectrum can be
shifted relative to the fixed cavity lines by tuning the laser frequency.
The frequency comb of the cavity modes is much denser than in the figure.}
\label{Fig:cavityFSR_comb}
\end{figure}

The cooling method we propose is based on the enhancement of the (anti-Stokes) Raman
transitions, where photons of an intense laser field are scattered by a molecular
dipole transition into the modes of a cavity thereby cooling the external and the
internal molecular motion, as sketched in Fig.~\ref{Fig:1}.
Spontaneous emission is largely suppressed, since the laser light is
far detuned from all direct electronic transitions.
The enhancement of coherent scattering into the cavity modes
is obtained by driving the Raman anti-Stokes transitions on resonance.
For this purpose, the laser frequency is varied sequentially,
in order to shift the comb of cavity resonances across the molecular anti-Stokes transitions,
as sketched in Fig.~\ref{Fig:cavityFSR_comb}.
The external motion is initially cooled to the cavity
linewidth by setting the laser frequency on the red-side of the Rayleigh
line~\cite{DomokosRitschJOSA03} in Fig.~\ref{Fig:cavityFSR_comb}. In our numerical simulations we checked
that this is a relatively fast stage, in which the internal motion
is not affected by the scattering processes. Then, the ro-vibrational
degrees of freedom are cooled as the laser frequency is sequentially
tuned to address all relevant anti-Stokes lines. The ro-vibrational
levels are hence sequentially emptied, till the molecules finally end up in their ground state.
%\end{widetext}

\section{Model and basic equations}
\label{model}

We now summarize the theoretical model, describing the quantum
dynamics of the molecular internal degrees of freedom and of the modes
of the cavity field.  We consider a gas
of molecules of mass $M$, prepared in the electronic ground state
$X$, and with dipole transitions $X\to {E}$. Here, ${E}=\left\lbrace A,...\right\rbrace$  is a set
of electronically excited states,
which may contribute significantly to the total polarizability. We
denote by $|\xi\rangle$, $|\xi^{\prime}\rangle$ the ro-vibrational states
of the electronic ground state, such that the state  $|X,\xi\rangle$ is at frequency
$\omega_{\xi}^{(X)}$, and by $|\xi''\rangle$ the ro-vibrational states
of the electronic excited states, with $|E,\xi''\rangle$ at frequency $\omega_{\xi''}^{(E)}$.
The elements of the dipole moment ${\bf d}$ are
\begin{equation}{\cal D}_{\xi\to\xi''}=\langle {E},\xi''|{\bf
d}|X,\xi\rangle.
\end{equation}
Denoting by ${\bf p}$, ${\bf x}$ the conjugate
momentum and position of the center of mass, the Hamiltonian
describing the dynamics of the uncoupled molecular degrees of
freedom reads
\begin{eqnarray}
H_{\rm mol}
&=&\frac{{\bf
p}^2}{2M}+\sum_{\xi}\hbar\omega_{\xi}^{(X)}\ke{X,\xi}\langle
X,\xi|\\
& &+\sum_{E,\xi''}\hbar\omega_{\xi''}^{(E)}\ke{E,\xi''}\langle E,\xi''|.\nonumber
\end{eqnarray}
The transitions between molecular levels are driven by a far-off
resonant laser and interact with an optical resonator as
illustrated in Fig.~\ref{Fig:1}a), undergoing a coherent dynamics
which is described by the total Hamiltonian $H=H_{\rm
mol}+H_c+V_c+V_L$. Here,
\begin{equation}
H_{\rm c}=\sum_c\hbar\Omega_c a_c\da a_c
\end{equation}
is the
Hamiltonian for the cavity modes, where $a_c$ ($a_c^{\dagger}$)
denotes the annihilation (creation) of a photon with energy
$\hbar\Omega_c$, with wave vector ${\bf k_c}$, polarization $\epsilon_0$ and the
(vacuum) electric field amplitude ${\cal E}_c$.
The coupling between cavity modes and molecular dipole transitions
is described by the term
\begin{eqnarray}
V_{\rm c}=\sum_{c,\xi',E,\xi''}\hbar
g_{c,\xi'\to\xi''}(z)\hat{a}_c\left({\hat{\sigma}_{\xi',\xi''}}^{E\dagger}+\hat{\sigma}^E_{\xi',\xi''}\right)+{\rm
h.c.},
\end{eqnarray}
where $g_{c,\xi'\to\xi''}(z)={\cal
E}_c(z)(\epsilon_0\cdot{\cal D}_{\xi' \rightarrow \xi''})/\hbar$ is the
vacuum Rabi coupling, whose dependence on the position along the cavity $z$ axis
is determined by the corresponding spatial mode function, and
\begin{eqnarray}
\hat{\sigma}^E_{\xi,\xi''}=\ke{X,\xi}\langle E,\xi''|, ~~~
\hat{\sigma}_{\xi,\xi''}^{E\dagger}=\ke{E,\xi''}\langle X,\xi|
\end{eqnarray}
are the corresponding dipole lowering and rising
operators. The molecules are pumped by a laser, which couples to the
dipole and is described by a classical field with amplitude ${\cal
E}_L$, polarization $\epsilon_L$, frequency $\omega_L$, and wave
vector ${\bf k_L}$. The laser is assumed to propagate
perpendicular to the cavity axis, and the Hamiltonian term
describing the coupling with the dipole takes the form
\begin{equation}
V_L=\sum_{\xi,E,\xi''}\hbar\Omega_{L,\xi\to\xi''}{\rm
e}^{-{\rm i}(\omega_Lt-{\bf k_L}\cdot{\bf
x})}\left(\hat{\sigma}_{\xi,\xi''}^{E\dagger}+\hat{\sigma}^E_{\xi,\xi''}\right)+{\rm
h.c.},
\end{equation}
where $\Omega_{L,\xi\to\xi''}={\cal
E}_L({\cal D}_{\xi\to\xi''}\cdot \epsilon_L)/\hbar$ gives the
corresponding coupling strength. In the interaction Hamiltonian $V_L$, we
keep the counterrotating terms, as we will consider parameter
regimes where the rotating wave approximation is not valid.

Finally, cavity modes and dipole transitions couple to the modes
of the electromagnetic field external to the resonator. We denote
by $\Gamma_{\xi'' \to \xi'}$ the decay rate along the dipolar
transition $|E,\xi''\rangle\to|X,\xi'\rangle$, such that the
linewidth of the state $|E,\xi''\rangle$ is
$\Gamma_{\xi''}=\sum_{\xi'}\Gamma_{\xi''\to\xi'}$. The linewidth
of the resonator modes is $2\kappa$, and we assume the regime in
which the Free Spectral Range (FSR) $\Omega_{c+1}-\Omega_c\gg
2\kappa$.

We assume that the laser and the cavity modes are far-off
resonance from the molecular transitions, as depicted in Fig.~\ref{Fig:1}b),
such that the
most relevant processes are Raman transitions, i.e., incoherent
Raman scattering in which a laser photon is absorbed and then
emitted spontaneously, and coherent Raman scattering in which a
laser photon is scattered into one cavity mode. The rate of
spontaneous Raman scattering reads
\begin{equation}
\label{eq:GammaG}
\Gamma_{\xi\to\xi'}^{\gamma}({\bf p})
=\sum_{E,\xi''}\Gamma_{\xi''\to\xi'}
   \left(\gamma_{E,\xi\to\xi''}^+({\bf
p}) +\gamma_{E,\xi\to\xi''}^-({\bf p})\right)
\end{equation}
and it
depends on the molecular momentum ${\bf p}$ through the Doppler
effect as
\begin{eqnarray}
\label{eq:Gammag}
\gamma^+_{E,\xi\to\xi''}({\bf p})&=&\frac{
    \Omega_{L,\xi\to\xi''}^2}{(\Delta_{\xi,\xi''}^E+{\bf k_L}\cdot {\bf
    p}/M)^2+\Gamma_{\xi''}^2/4}\\
\gamma^-_{E,\xi\to\xi''}({\bf p})&=&\frac{
    \Omega_{L,\xi\to\xi''}^2}{(\Delta_{\xi,\xi''}^E-2\omega_L-{\bf k_L}\cdot {\bf
    p}/M)^2+\Gamma_{\xi''}^2/4},
\end{eqnarray}
where
$\Delta_{\xi,\xi''}^E=\omega_{\xi}^{(X)}-\omega_{\xi''}^{(E)}+\omega_L$
denotes the detuning between laser and internal transition. The rate $\Gamma_{\xi\to\xi'}^{\kappa}({\bf p})$ gives the
scattering rate of a photon from the laser into the cavity mode
and its subsequent loss from the cavity.

To evaluate the rate of photon scattering into the cavity modes,
we consider a standing wave cavity, and focus onto the regime in which the cavity photon is not reabsorbed
by the molecules but lost via cavity decay~\cite{Vuletic00}.
Moreover, we assume that the molecular kinetic energy exceeds
the height of the cavity potential. In this regime, we decompose the coupling constant
$g_{c,\xi'\to\xi''}(z)$ into the Fourier components $g_{c,\xi'\to\xi''}^{\pm}$ at cavity-mode
wave vector $\pm |{\bf k_c}|$, and write the rate of photon scattering into the cavity as $\Gamma_{\xi\to\xi'}^{\kappa}({\bf
p})=\Gamma_{\xi\to\xi'}^{\kappa,+}({\bf p})+\Gamma_{\xi\to\xi'}^{\kappa,-}({\bf p})$, where the sign $\pm$
gives the direction of emission along the cavity axis and
\begin{eqnarray}
    &&\Gamma_{\xi\to\xi'}^{\kappa,\pm}({\bf p})
    =2\kappa\sum_{c,E,\xi''}\Bigl|g_{c,\xi''\to\xi'}^{\pm}\Bigr|^2\label{eq:GammaK}\\
    & &\left(\frac{\gamma^+_{E,\xi\to\xi''}({\bf
p}) }
    {\left(\delta\omega^+\pm
    {\bf k_c}\cdot {\bf p}/M\right)^2+\kappa^2}+\frac{\gamma^-_{E,\xi\to\xi''}({\bf
p})}
    {\left(\delta\omega^-\pm
    {\bf k_c}\cdot {\bf p}/M\right)^2+\kappa^2}\right),
    \nonumber
\end{eqnarray} where
$\delta\omega^\pm=\omega_{\xi}^{(X)}-\omega_{\xi'}^{(X)}\pm\omega_L-\Omega_c$
is the frequency difference between initial and final (internal
and cavity) states. We note that the reabsorption and
spontaneous emission of the cavity photon can be neglected when
\begin{equation}
\kappa\gg |g_{c,\xi'\to\xi''}\Omega_{L,\xi\to\xi''}/\Delta_{\xi,\xi''}^E|. 
\end{equation}
The basic condition for cavity cooling is that the rate of photon
scattering into the cavity exceeds the corresponding spontaneous
Raman scattering rate,
\begin{equation}
\Gamma_{\xi\to\xi'}^{\kappa}({\bf
p})\gg\Gamma_{\xi\to\xi'}^{\gamma}({\bf p}).
\end{equation}
This is the regime on which we will focus.

\subsection{Rate equations}

We can now write the
equations describing the damped dynamics of the molecules.
For this purpose, we denote by ${\cal W}_{\xi}(p)$ the occupation of the
molecular state $|X,\xi\rangle$ at momentum $p$.

An equation for the kinetic energy can be simply derived in the
semiclassical limit. Let us first define by $\Delta p^2/2M$ the
average change in kinetic energy in an infinitesimal
interval of time $\Delta t$. This is given by $\Delta p^2/2M=\sum_{\xi}\Delta
p_{\xi}^2/2M$, where
\begin{eqnarray}
\Delta p_{\xi}^2 &=&\Delta t\int {\rm d}p
\sum_{\xi',j=\pm}\\
& &\Bigl[(p+j\hbar
k)^2\Gamma_{\xi'\to\xi}^{j}(p){\cal W}_{\xi'}(p)-p^2\Gamma_{\xi\to\xi'}^{j}(p){\cal W}_{\xi}(p)\Bigr]\nonumber
\end{eqnarray} and we have introduced, for convenience, the
total rate $\Gamma_{\xi'\to\xi}^{\pm}(p)=\Gamma_{\xi'\to\xi}^{\kappa,\pm}(p)+\Gamma_{\xi'\to\xi}^{\gamma}(p)/2$.
In the regime where the laser is tuned such that only Rayleigh scattering into the cavity mode is enhanced,
the relevant processes will not change the internal state of the molecule. In this regime, when the rate of spontaneous Raman scattering is much smaller than the coherent cavity scattering one obtains the typical dynamics of Doppler cooling, see~\cite{Itano} for a detailed treatment. Here, the cooling rate is given by $(\hbar^2k^2/2M)\Gamma'(0)$, where $\Gamma'(0)\approx \sum_{\xi}\Gamma_{\xi\to\xi}^{j}(0)$
and the asymptotic value of the kinetic energy is
\begin{equation}
E_{\rm kin}^{\infty}=
\frac{\hbar}{4}\frac{(\omega_L-\Omega_c)^2+\kappa^2}{|\omega_L-\Omega_c|},
\end{equation}
which is minimum for $\omega_L-\Omega_c=-\kappa$
and takes the value
$E_{\rm kin}^{\infty,{\rm min}}=\hbar\kappa/2$ \cite{DomokosRitschJOSA03}.

The generic rate equation, describing the dynamics of the population of the
molecular state ${\cal P}_{\xi}=\int dp {\cal W}_{\xi}(p)$ reads
\begin{eqnarray} \dot{\cal P}_{\xi} &=&-\int
dp\sum_{\xi'}\left(\Gamma_{\xi\to\xi'}^{\kappa}(p)+\Gamma_{\xi\to\xi'}^{\gamma}(p)\right){\cal
W}_{\xi}(p)\nonumber\\
& &+\int
dp\sum_{\xi'}\left(\Gamma_{\xi'\to\xi}^{\kappa}(p)+\Gamma_{\xi'\to\xi}^{\gamma}(p)\right){\cal
W}_{\xi'}(p),\nonumber\\
\end{eqnarray}
which requires the knowledge of all rates and
populations at each instant of time. Assuming that the molecules
have been previously cavity cooled to the cavity linewidth, and
that the momentum distribution is not significantly modified by the
inelastic processes that change the internal state of the
molecule, then the rate of inelastic processes is
\begin{equation}
\int dp
\Gamma_{\xi\to\xi'}^{\kappa,\pm}(p){\cal
W}_{\xi}(p)\approx\Gamma_{\xi\to\xi'}^{\kappa,\pm}(0){\cal
P}_{\xi}.
\end{equation}
Hence, in this regime, the incoherent dynamics can be
described by the rate equation
\begin{eqnarray}
\dot{\cal P}_{\xi}
&=&-\sum_{\xi'}\left( \Gamma_{\xi\to\xi'}^{\kappa}+\Gamma_{\xi\to\xi'}^{\gamma}\right){\cal
P}_{\xi}\\
&
&+\sum_{\xi'}\left(\Gamma_{\xi'\to\xi}^{\kappa}+\Gamma_{\xi'\to\xi}^{\gamma}\right){\cal
P}_{\xi'},\nonumber
\end{eqnarray} whose stationary solution is found using
the detailed balance principle.

The simulation of the cooling procedure uses rate equations for the molecular
levels with the rates~(\ref{eq:GammaG}) and~(\ref{eq:GammaK}) as evaluated in the previous section.
The sequence for the laser driven transitions
is simulated by changing the detuning at the different steps.
The cooling of the external degrees of freedom is the preliminary step.
We have evaluated the cooling rate for the external motion with the laser
frequency set to $\omega_L=\Omega_c-\kappa/2$, and checked that the internal molecular
level distribution is practically unaffected during the cooling time.
In the step for cooling the internal degrees of freedom, we then assumed
that the motion was cooled at the Doppler limit, and set $p=0$ in
the rates ~(\ref{eq:GammaG}) and~(\ref{eq:GammaK}).
In the next section we report the details of the numerical treatment,
for determining the parameters which give us the rates.

\section{The molecular systems: OH and NO}
%\section{Application to molecules}
\label{QChemistry}

We simulate the cooling scheme for OH and NO radicals.
Cold ensembles of these two molecules are experimentally available
\cite{Chandler03,vdMeerakker05,Bochinski04,Fulton06}.
We first apply the cooling scheme for an
internal temperature of $300$\,K. Since at this temperature OH
molecules are in the vibrational ground state,
we also simulate the process for OH in a situation in which initially
the internal temperature is artificially
high, in order to investigate cooling of vibrational motion explicitly.

OH and NO can be modeled accurately by \emph{ab initio}
quantum chemistry methods. They have a similar electronic
valence structure which allows to compare their electronic properties.
In particular, the unpaired electron leads to a significant spin orbit
splitting of $139$\,cm$^{-1}$ and $123$\,cm$^{-1}$ for the $X^2\Pi$
ground states of OH and NO, respectively. $\Lambda$-doubling
occurs due to the coupling between rotation and electronic angular momentum.
The doubling lies in the MHz
regime and is different for the $X^2\Pi_{3/2}$
and the $X^2\Pi_{1/2}$ components. We only consider one
$\Lambda$-state, since states of different parities are not coupled
by Raman transitions. The hyperfine splittings in OH, due to the nuclear
spin of the hydrogen atom, are neglected since angular momentum conservation
for the rotational Raman transitions inhibits transitions between these sublevels.
Hence, we assume that it is possible to prepare the molecules in the lower lying
$X^2\Pi_{3/2}$ state for OH and $X^2\Pi_{1/2}$ for NO.
With these assumptions we choose $J$ as the rotational quantum
number. Because of the Raman selection rule $\Delta J = 0,\pm 2$, matrix elements between
rotational states with even and odd $J$ values vanish.
This leads to two separate ladders for the scattering process with ground states
$\ke{X,v=0,J=0}$ and $\ke{X,v=0,J=1}$, respectively.

The energy of the rotational levels reads
\begin{equation}
E_J = B(v)J(J+1)-D_j J^2(J+1)^2 ,
\end{equation}
with
\begin{equation}
B(v) = B_e + B_{e,x1} (v+\frac{1}{2})+B_{e,x2} (v+\frac{1}{2})^2 .
\end{equation}
The constants, including the anharmonicities, are taken from Ref.~\cite{Huber77}. In particular, for OH
they read$(B_e,$ $B_{e,x1},$ $B_{e,x2},$ $\omega_e) =$ $($18.871, -0.714, 0.0035, 3735.21$) \,{\rm cm}^{-1}$
while for NO they are
$(B_e, B_{e,x1}, B_{e,x2}, \omega_e) =$ $($1.7042, -0.01728, 0.000037, 1904.2$) \,{\rm cm}^{-1}.$
The anharmonicity in the vibrational ladder is
included in the \emph{ab initio} potential energy surfaces (PES).

\subsection{Molecular polarizabilities}

We now determine the molecular rotational and vibrational structure
in order to evaluate the scattering rates for simulating the cooling dynamics. 
A relevant quantity is the polarizability $\alpha$, which is defined for molecular systems as
\begin{eqnarray}
\vect{\alpha} =
\sum_{E,\xi''} &\dfrac{ {\cal D_{\xi \rightarrow \xi''}} {\cal D_{\xi'' \rightarrow \xi}}}
  { \hbar \Delta_{\xi,\xi''}^{E}}+\dfrac{ {\cal D_{\xi \rightarrow \xi''}} {\cal D_{\xi'' \rightarrow \xi}}}
  {\hbar (\Delta_{\xi,\xi''}^{E} - 2 \omega_L)} ,
\label{eq:pol}
\end{eqnarray}
and is expressed as a function of the laser frequency $\omega_L$ and the internal nuclear
coordinates. Note that also higer lying excited states can have
significant contributions.
The polarizability is described as a tensor which can be expressed
in terms of irreducible components.
The isotropic part $\alpha^{(0)}$, defined as
\begin{equation}
\alpha^{(0)} = \frac{1}{3} Tr(\vect{\alpha}),
\end{equation}
is responsible for transitions which do not change the angular momentum of the molecule.
The traceless component $\alpha^{(2)}_{zz}$ along the main
axis of the molecule,
can be evaluated from
\begin{equation}
\label{eq:alpha_trl}
\alpha^{(2)}_{ij} = \alpha^{(2)}_{ji} = \frac{1}{2}
    \left( \alpha_{ij} + \alpha_{ji} \right) - \alpha^{(0)}.
\end{equation}
This component is needed to describe ro-vibrational transitions
associated with a change in $J$. In all subsequent formulas
$\alpha$ represents the appropriate component.
This general formulation allows us to treat all possible
transitions between the internal states.

The transitions between the vibrational states
can be written as the matrix element
\begin{equation}
\alpha_{v \rightarrow {v'}} =\br{\xi_v} \alpha(R) \ke{\xi_{v'}} ,
\end{equation}
in which the integration is over
the nuclear coordinates $R$ and where
$\xi_{v}$ indicates the pure vibrational states.
The transition strength between the rotational
levels is evaluated using the Placzek-Teller
coefficients \cite{placzekteller_org,placzekteller}
for general diatomic molecules (shown in Tab. \ref{tab:PlaczekTeller}).
The Placzek-Teller coefficient  $S_{J\rightarrow J'}$
for the corresponding transition is
multiplied with the square of the absolute value of the polarizability,
giving the complete transition matrix element
\begin{equation}
| \alpha_{v,J \rightarrow v',J'} |^2= | \alpha_{v \rightarrow {v'}} |^2 \cdot S_{J \rightarrow{J'}} .
\end{equation}
The obtained value for $\alpha$ can be inserted in Eq.~\ref{eq:GammaG}
and Eq.~\ref{eq:GammaK} to calculate the transition rates in
the molecular system.

\begin{table}[h]
\caption{Placzek-Teller coefficients $S$ adapted to diatomic molecules.
With the different cases for $J$ one can calculate Stokes and anti-Stokes
transitions as well as Rayleigh scattering.}
\label{tab:PlaczekTeller}
\begin{center}
\begin{tabular}{c|c}
$\Delta J$ & $S$ \\ \hline
&  \\
$J$ & $\dfrac{\left[ J(J+1)\right] ^2}{J(J+1)(2J-1)(2J+3)}$\\
 & \\
$J+2$ & $\dfrac{3 (J+1)^2 (J+2)^2}{2(J+1)(J+2)(2J+1)(2J+3)}$\\
 & \\
$J-2$ & $\dfrac{3 (J-1)^2 J^2}{2(J-1)J(2J+1)(2J-1)}$ \\
 & \\
\end{tabular}
\end{center}
\end{table}
\begin{figure}[h]
\includegraphics{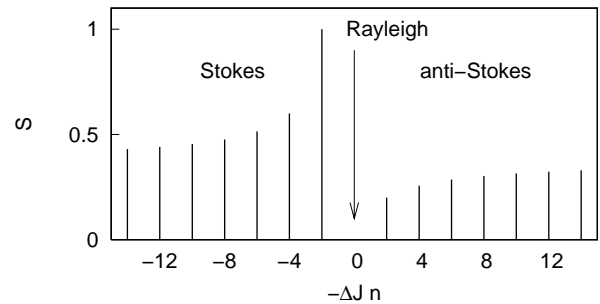}
\caption{Illustration of the Placzek-Teller coefficients $S$ for the
seven relevant Raman lines. The transitions indexed with $n$, which are
allowed according
to the selection rule $\Delta J = \pm 2$ are shown.
For example, $-\Delta J n=2$ refers to $J_{2 \rightarrow 0}$ and
$-\Delta J n=4$ refers to $J_{3 \rightarrow 1}$.
The coefficients
for the Rayleigh scattering $\Delta J = 0$ are not included in the
picture.}
\label{Fig:placzek-teller}
\end{figure}
\begin{figure}[h]
\includegraphics{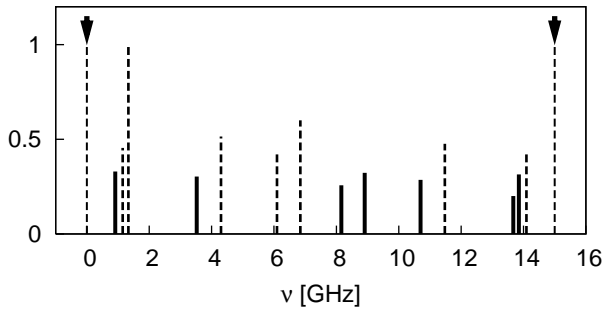}
\caption{Reduced spectrum of the seven relevant Raman lines for OH.
The Raman transitions are projected into one FSR. The solid lines
indicate the driven anti-Stokes transitions. The dashed lines represent the
corresponding Stokes lines in the Spectrum. The cavity resonances are the lines
on left and on the right side highlighted by the arrows.}
\label{Fig:modFSR}
\end{figure}

\subsection{{\it ab initio} calculations}

We carried out
different types of \emph{ab initio} calculations
for all the relevant properties not directly
available from literature.
The PES were
obtained with Molpro \cite{molpro} in
a complete active space self consistent field
(CASSCF) calculation and were further improved with a
multi reference configuration interaction (MRCI)
step. A valence triple zeta single atom basis set
(cc-pVTZ) \cite{cc-pVXZ} was used. For the OH radical we chose
an active space of 11 orbitals which
includes all electrons except
the 1s core orbital from the oxygen.
The NO was treated in a similar way with
12 active orbitals and only leaving out the
1s core orbitals from the oxygen and the
nitrogen atom.

For the calculation of the polarizability
linear response theory on the same CASSCF \cite{linres1,linres2}
level were performed. This
method is implemented in the program package
DALTON \cite{dalton}.
The quality and convergence of the
resulting polarizabilities were
checked through several tests.
The finite field method is used for
comparison with the values in the
static limit where $\omega_L = 0$.
The static polarizability of a particle
can expanded in a Taylor series
with respect to the perturbation of an
electric field, such that $\alpha$ is found from
\begin{equation}
\alpha^{FF} = \frac{\partial^2E(\vec{\cal E})}{\partial{\cal E}_i \partial{\cal E}_j} .
\end{equation}
The differentiation is done numerically with
small field strengths.
With this method several basis set checks (see Tab. \ref{tab:BS_OH})
were performed for OH on the CASSCF level mentioned
before.
\begin{table}
\caption{Single point polarizability calculations
for OH with various basis sets. All calculations
were carried out with on the CASSCF level of theory
with the finite field method (${\cal E} = 0.001$\,au)
at the equilibrium geometry.
Listed are the needed tensor components $\alpha^{(0)}$ and $\alpha^{(2)}_{zz}$.
Additionally the corresponding dipole moment is printed
(The experimental value is $\mu_z = 0.657$\,au$=1.67$\,Debye \cite{Dipmom_OH}).
All values are given in atomic units.}
\begin{center}
\begin{tabular}{l|rrrrrr}
Basis set   & $\alpha^{(0)}$& $\alpha_{zz}^{(2)}$ & $ \mu_z$ \\ \hline
6-31G(d,p)  & 3.809 & 2.172 & 0.711 \\
6-31G++(d,p)    & 4.936 & 1.652 & 0.741 \\
aug-cc-pVDZ & 6.826 & 1.438 & 0.648 \\
cc-pVTZ(d,p)    & 5.194 & 1.836 & 0.667 \\
cc-pVTZ     & 5.105 & 1.928 & 0.668 \\
aug-cc-pVTZ(d,p)& 7.230 & 1.188 & 0.648 \\
aug-cc-pVTZ & 7.240 & 1.200 & 0.645 \\
aug-cc-pVQZ & 7.407 & 1.069 & 0.646 \\
aug-cc-pVQZ(d,p)& 7.403 & 1.060 & 0.647 \\
\end{tabular}
\label{tab:BS_OH}
\end{center}
\end{table}
It turned out (as already mentioned in earlier
publications \cite{CH2O_alpha}) that the results are very sensitive to the
choice of the basis set. The addition of diffuse
functions gives the better results.
Higher angular momentum f-polarization
functions have only a little effect on the polarizability.
It turned out that correlation consistent basis sets (cc-pVXZ)
\cite{cc-pVXZ,cc-pVXZ-aug}
are more adequate for this task than the widely
used 6-31G basis sets \cite{631G}.
The aug-cc-pVTZ basis set was chosen for
the polarizability calculations as a good
compromise between size and quality of the result.
This basis set was used to compare the result
from the finite field calculations against
linear response calculation in the static
limit at zero frequency, where both methods should give equal
results (Tab. \ref{tab:methods_OH}). The
obtained values differ less than $1$\,\% and
hence we can assume to have reasonable values.
\begin{table}
\caption{Comparison between the OH polarizabilities
from the finite field and the linear response
calculations at the CASSCF/aug-cc-pVTZ(d,p) level of theory.
Additionally the dynamic polarizability at
$532$\,nm is shown.
All values are given in atomic units.}
\begin{center}
\begin{tabular}{l|rrrrrr}
Method   & $\alpha^{(0)}$ & $\alpha_{zz}^{(2)}$\\ \hline
Finite field        & 7.23 & 1.19\\
Linear response, static limit   & 7.21  & 1.17\\
Linear response, $\lambda = 532$\,nm    & 7.39  & 1.15\\
\end{tabular}
\label{tab:methods_OH}
\end{center}
\end{table}
\begin{figure}[t]
\includegraphics{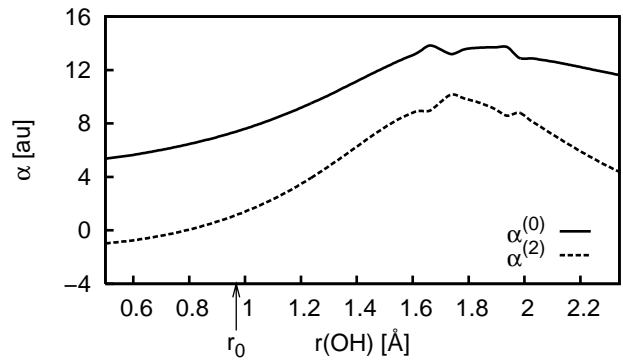}
\caption{The polarizability function in dependence of the bond length
for OH at a laser wavelength of 532\,nm. The isotropic part $\alpha^{(0)}$ is used
for the calculation of the Rayleigh scattering transitions. The traceless part
$\alpha_{zz}^{(2)}$ of the tensor determines the rotational transitions. The equilibrium
geometry is indicated by $r_0$. The wiggles between
$1.6$\,\AA{} and $2.0$\,\AA{} are artifacts from resonances where the detuning
$\Delta^E_{\xi,\xi''}$ is near zero.
These singularities were smoothed by an interpolation.}
\label{Fig:OH_pol_func}
\end{figure}
The calculated polarizability function in dependence
of the OH bond length is shown in Fig. \ref{Fig:OH_pol_func}.
The function has been calculated for a laser
wavelength of $532$\,nm. Both curves for
the isotropic and the traceless part show a
typical increase with greater bond lengths
reaching a maximum between $1.6$\,\AA{}
and $2.0$\,\AA{}. In this area the $X^2\Pi$ and
$A^2\Sigma^+$ state come closer	
(see Fig. \ref{Fig:1}) and their energy difference
matches the laser frequency.
Here, the Raman approximation \cite{Shore} becomes
invalid and the evaluated data points show a
singularity. This effect is smoothed by
an interpolation routine.
The remaining wiggles have no effect on the results as they lie
outside the nuclear range relevant in the considered Raman process.

The vibrational eigenvalues and eigenfunctions
of the molecules were evaluated with a
relaxation method using propagation in imaginary
time with an additional diagonalization step \cite{Sundermann}.
This method is used to determine the vibrational
wavefunctions in the ground state and
first electronic excited state, necessary to obtain
the relevant Franck-Condon factor of the system.

\section{Experimental parameters}
\label{Experimental}

We now discuss the experimental parameters, which are required in order to implement
efficiently the cooling scheme. We
assume that the OH or NO molecules are prepared in the lower-lying
component of the $X^2\Pi$ electronic ground state
with a motional temperature around or slightly below 1\,K. This
could be realized with, e.g., helium-buffer-gas
cooling~\cite{Weinstein98}, electrostatic
filtering~\cite{RangwalaPRA03}, or decelerator
techniques~\cite{BethlemPRL99,Barker}. However, as the purpose of
this paper is to demonstrate in particular the cooling of the
internal motion of the molecule, the filtering
method~\cite{RangwalaPRA03} matches most closely our assumed
initial conditions. At $T\approx 300$\,K only the first nine
rotational levels of OH are occupied
with a population $> 10^{-3}$. The first vibronic excitation
for OH has an energy of $3727.95$\,cm$^{-1}$ \cite{Huber77},
and thus only
the ground state $v=0$ is occupied. For NO
there are 25 rotational levels and because the
vibronic excitation lies at $1904.2$\,cm$^{-1}$
again only the ground state is occupied.

We choose a laser wavelength of 532\,nm. Since this can be
generated by frequency doubling of 1064\,nm light, commercial
systems are available with ample power, with high quality of the
beam profile and which run at single frequency. The frequency of
this light is far below that of the OH A-X ro-vibronic band.
We assume to have single-frequency light of 10\,W
enhanced by a factor of 100 by a build-up cavity in a TEM$_{00}$
mode, corresponding to a Rabi coupling $\Omega_{L,0\to 0}=2
\pi\times 69\,$GHz and frequency
$\omega_L=\omega_0^{(A)}-\omega_0^{(X)}-\Delta$ with
$\Delta\approx 2\pi\times407\,$THz. Note that
these intracavity intensities are routinely achieved in some
commercial devices.

The laser frequency should be varied during cooling, in order to
drive (quasi-)resonantly the cooling transitions. In combination
with the broad spectrum of cavity modes, the laser only needs to
be varied over one FSR to address all anti-Stokes lines.
Fig.~\ref{Fig:modFSR} displays the Stokes and anti-Stokes Raman lines as a
function of the frequency modulus the FSR. Most commercial systems
delivering high-power 532\,nm light are not tunable, but this does
not seem a fundamental problem. The build-up cavity will require
tracking while the pump laser is scanned, which is also possible
with today's technology.

We choose a high-finesse optical Fabry-Perot-type cavity with a
length of $L=1\,$cm, which seems large enough to accommodate a
trapping device as e.g., demonstrated in refs.~\cite{Drewsen04,Schiller05,BethlemNature00,RiegerPRL05}.
This length leads to a free-spectral range (FSR) of $15\times2\pi\,$GHz. For simplicity
we assume that the cavity only supports zeroth-order transverse
modes. The cavity half-linewidth is set to
$\kappa=75\times2\pi\,$kHz, and the coupling $g_{c,0\to 0} = 2
\pi\times 116$ kHz. This is achieved with a mode volume of $3.2
\times 10^{-13}\,$m$^{3}$, assuming a mode waist of $w_0 = 6\,\mu$m,
and a cavity finesse $F=10^5$, i.e., a mirror reflectivity of
$0.999969$. In practice it might be necessary to choose a
near-degenerate setup, e.g. a near-confocal cavity, in order to
avoid higher-order cavity modes from appearing everywhere in the
reduced spectrum, possibly coinciding with Stokes lines.
Choosing such a degenerate setup actually increases the
possibilities of scattering light into the cavity, increasing the
cooling rate. There is one caveat: if the higher-order transverse
cavity modes are not exactly degenerate with the zeroth order,
this will lead to a cavity linewidth which will appear broadened
and all lines in the reduced spectrum should be convoluted with
this cavity linewidth. However, we do not expect this to be an
issue, not even in a near-degenerate cavity, as our scheme is
robust against a small number of coincidences between Stokes and
anti-Stokes lines.

\section{Cooling simulation results}
\label{Results}

\subsection{Rotational cooling of OH and NO radicals}

\begin{figure}[h]
\includegraphics{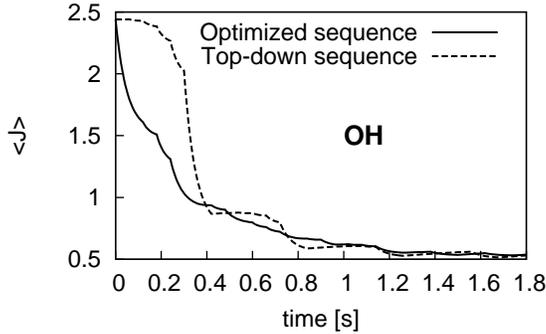}
\caption{Rotational cooling of OH starting from 300\,K. The expectation value
$\left\langle J \right\rangle=\sum_J {\cal P}_J J$ of the rotational quantum number
is plotted against time. The solid line shows a cooling process
with a manually optimized laser sequence. The dashed line represents
a ``top-down`` process where all higher levels are emptied sequentially in multiple cycles.}
\label{Fig:OH_meanJ}
\end{figure}

\begin{figure}[h]
\includegraphics{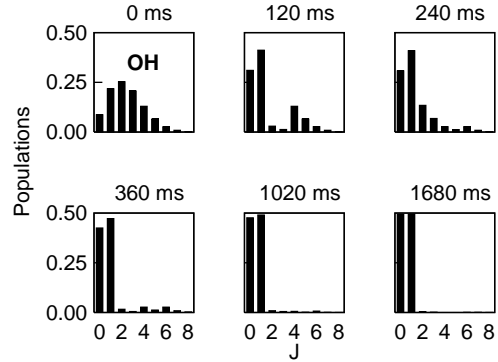}
\caption{Populations at selected times of the cooling process for
the manually optimized sequence. Each transition is driven for $60$\,ms.}
\label{Fig:OH_pops}
\end{figure}

We now simulate the cooling of OH and NO with the experimental parameters as described above.
We start with the OH radical as the
molecule of main interest.
Fig. \ref{Fig:OH_meanJ} and Fig. \ref{Fig:OH_pops} show the results of the
simulation.
The FSR has been fine tuned so that the transitions
$J_{3 \rightarrow 1}$  and $J_{2 \rightarrow 0}$ could
be pumped simultaneously.
In general, the spectral lines of a molecular system
are not equally spaced and in general only one transition at a time
can be addressed.
However, two transitions can be selected simultaneously
by tuning the FSR so that
two cavity lines coincide with two transitions.

In Fig. \ref{Fig:OH_meanJ} two different sequences
are shown. The dashed curve represents a simple ``top-down`` variant
which cools down the molecule to the ground states beginning from the
highest occupied level. The solid curve is  a
manually optimized sequence in order to obtain a
faster decrease of the expectation value $\left\langle J \right\rangle=\sum_J {\cal P}_J J$.

\begin{figure}[h]
\includegraphics{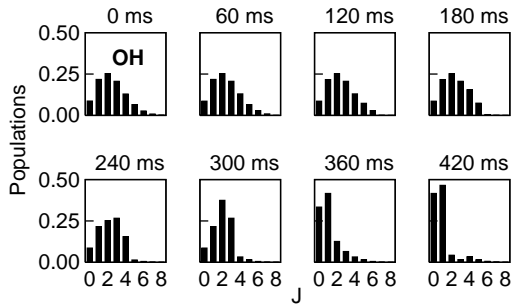}
\caption{Populations at each step of the cooling process for the
``top-down`` sequence in the first cycle. Each transition is driven for $60$\,ms.}
\label{Fig:OH_pops_td}
\end{figure}

Snapshots of the population time evolution are shown
in Fig.~\ref{Fig:OH_pops_td} and Fig.~\ref{Fig:OH_pops}.
In the top-down sequence, Fig. \ref{Fig:OH_pops_td}, we start driving the
transitions
corresponding to the highest occupied states, $J_{8\rightarrow 6}$ and
$J_{7\rightarrow 5}$, continuing with
the subsequent anti-Stokes transitions.
Every step lasts about $60$\,ms, whereby this optimal value was found empirically.
After the first cycle $88$\,\% of the population is
in the states $J=0,1$. To achieve a lower final temperature
the procedure is repeated six times.
The rate for the decrease in $\left\langle J \right\rangle$ is $2.3$\,Hz.
In the optimized sequence, first the $J=2$ and $J=3$ levels are addressed. In Fig. \ref{Fig:OH_pops} this can be seen in the snapshot at $120$\,ms. Here a large population fraction is transferred to the two ground states. In the next step
(after $240$\,ms) the transitions $J_{5 \rightarrow 3}$
and $J_{4 \rightarrow 2} $ are driven sequentially for $60$\,ms each.
Repeating the first step, more
than $90$\,\% of the population is transferred to the
ground states. By addressing
the levels $J>5$ and repeating the whole
procedure, $98.8$\,\% of the population is transferred to the two ground states $J=0$ and $J=1$ after $1680$\,ms.
The achieved rate for the decrease in $\left\langle J \right\rangle$ is $3.6$\,Hz.

We also checked the effect of spontaneous Raman-processes
on the vibrational state distribution when the rotational motion is cooled, and found
only little vibrational heating.
After a complete sequence of $1.8$\,s less than $1$\,\% is in the state $v=1$.
Without cavity cooling,
the rotations are heated, such that,
the expectation value $\left\langle J \right\rangle$ rises from $2.44$ to $2.56$
on time scales of the order of $1.8$\,s.

\begin{figure}[h]
\includegraphics{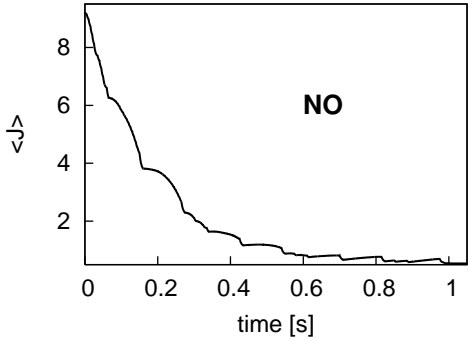}
\caption{Rotational cooling of NO with an initial temperature of $300$\,K.
The sequence for driving the transitions was manually optimized.
Due to the larger polarizability the NO radical can be cooled
faster than OH although more states are occupied.}
\label{Fig:NO_meanJ_300K}
\end{figure}

\begin{figure}[h]
\includegraphics{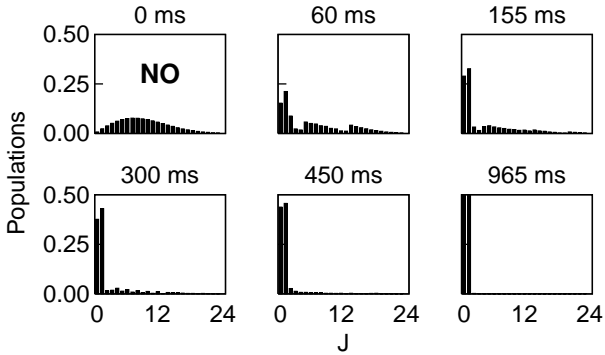}
\caption{Populations during the cooling process for NO with an optimized sequence.
The initial distribution corresponds to a temperature of $300$\,K.
Various stages of the cooling process are shown.}
\label{Fig:NO_pops}
\end{figure}

To analyze the dependence of the cooling process
on the related molecular properties,
we now simulate cooling of a NO molecule, whose level structure in the electronic ground state
is similar to the OH radical while the rotational and
vibrational constants differ. In contrast to the more elaborate calculations of OH,
the polarizability of NO was calculated in the quasi-static limit.
This is a reasonable approximation
as long as the ground state is coupled far-off resonance from the excited state.
Due to the
$\pi$-bonding in the NO radical the
value for the polarizability is
about 4 times greater than in the OH radical. Hence, compared to OH,
NO offers  a larger polarizability, which increases the cooling rate.
On the other hand, due to the smaller rotational constant 25 rotational states
are occupied at 300\,K. Thus, NO is characterized by a broader distribution
over the rotational levels than OH, which increases the cooling time.
The result for the cooling
procedure is shown in Fig. \ref{Fig:NO_meanJ_300K}
and Fig. \ref{Fig:NO_pops} for the manually-optimized sequence.
It turned out that it is a good choice to start the sequence by addressing
the levels near the maximum of the
initial Boltzmann distribution (Fig. \ref{Fig:NO_pops}),
in this case at levels $J=12,13$.
Every transition is driven for $5$\,ms.
After $60$\,ms the maximum of the distribution function is shifted to
the ground-state levels $J=0,1$. The next step is started in the states
$J=20,19$ and after $155$\,ms more than $50$\,\% are in the two
defined ground states. After $965$\,ms $99.2$\,\% of the
population is in the lowest two states. The optimal rate for the decrease in
$\left\langle J \right\rangle$ is $5.3$\,Hz.
In order to make a comparison with OH, we repeat the simulation for
NO by taking the same initial internal distribution
of only nine levels (which would correspond, in an experiment, to a sample of NO molecules at about $25$\,K).
Due to the larger polarizability of NO, the cooling is about 12 times faster than
in the case of OH, and the optimal rate for the decrease in $\left\langle J \right\rangle$ is $43$\,Hz.

\subsection{Cooling of the vibrational and rotational motion}

To test the efficiency of our scheme for cooling vibrational excitations,
we consider again OH molecules and
construct a scenario where the vibrational temperature is artifically high,
taking an initial distribution in which the first nine {\em vibrational} levels are
appreciably occupied. First we just cool the vibrational degrees of freedom,
neglecting the Placzek-Teller coefficients. In this case,
not only the absolute value of the polarizability, but also its finite slope
with respect to the internuclear coordinate (see Fig.\ref{Fig:OH_pol_func}) is important.

\begin{figure}
\includegraphics{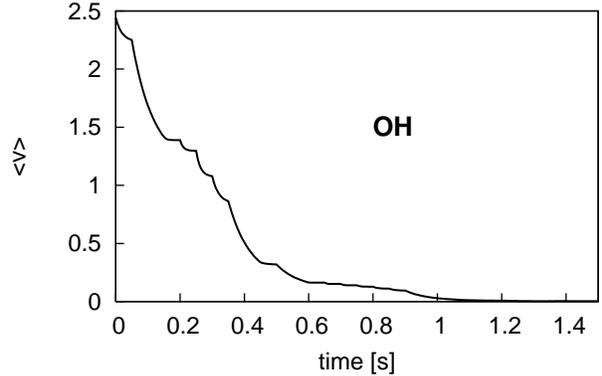}
\caption{Cooling of the vibrational motion of OH for an initial
thermal distribution where the first nine levels are significantly
occupied.}
\label{Fig:OH_vib}
\end{figure}

A roughly optimized frequency sequence for this procedure
leads to the fast cooling process shown in Fig. \ref{Fig:OH_vib},
demonstrating that the concept works. However,
it overestimates the cooling rate because the neglected Placzek-Teller coefficients for
anti-Stokes transitions are smaller than $0.5$. 

In the next step the rotational degrees of
freedom are combined with two vibrational levels
each containing nine rotational substates.
All initial population is in $v=1$, with a $300$\,K rotational temperature.
The task of achieving a good result for both
$\left\langle v \right\rangle$ and $\left\langle J \right\rangle$
is now more complicated but still feasible.
\begin{figure}[h]
\includegraphics{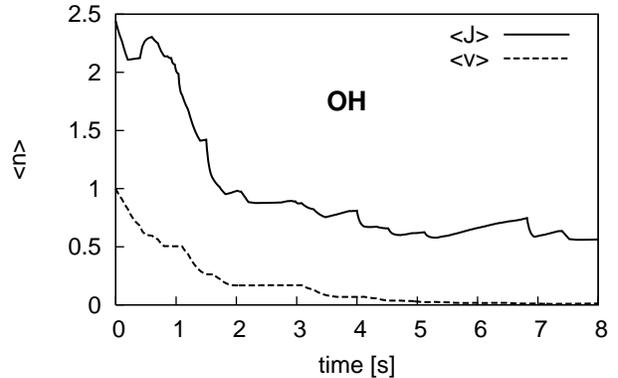}
\caption{The expectation value of $n$ as function of time,
where $n$ is a substitute for the quantum numbers $v$ and $J$, respectively.
All population is initially in vibrational state $v=1$ and the
rotational levels follow a thermal distribution at $300$\,K.}
\label{Fig:OH_vib_rot}
\end{figure}
The roughly optimized sequence of Fig.\ref{Fig:OH_vib_rot} is an example
where vibrational and rotational degrees of freedom have been cooled simultaneously.
For these results, the selection rules are exploited such as to reduce
$\left\langle v \right\rangle$ and $\left\langle J \right\rangle$
simultaneously as much as possible. However, as can be seen in the figure, this is not always possible.
In the beginning the transitions $v_{1\rightarrow 0}J_{3\rightarrow 1}$
and $v_{1\rightarrow 0}J_{2\rightarrow 0}$ are driven for $100$\,ms each.
In the next step the transition
$v_{1\rightarrow 0}J_{1\rightarrow 1}$ is driven.
Here $\left\langle v \right\rangle$ still becomes
smaller but $\left\langle J \right\rangle$ stays
constant. In the following step after, $300$\,ms, a compromise was made. The
state $v=1,J=0$ has to be emptied but the
logical target state $v=0,J=0$ is not accessible
because the Placzek-Teller coefficient for the
$J_{0\rightarrow 0}$ transition is zero. Therefore the next
accessible level, $v=0,J=2$, is used, which results
in an increase in $\left\langle J \right\rangle$.
Note that $J$ changes by $2$ and $v$ by $1$ and
thus the variations in $\left\langle J \right\rangle$
are larger than those in $\left\langle v \right\rangle$.
The further progression for the laser induced transitions
implies the alternation of two principles:
transferring populations to
the lower vibrational states like for
the first cooling steps and cooling the
rotational levels without vibrational heating.
It turned out that for the further cooling steps it is
more effective to first cool the rotational motion
without changing the vibrational motion before
transferring the population to the vibrational ground state.

In the end of the procedure $\left\langle v \right\rangle$
and $\left\langle J \right\rangle$
converge to their respective minima $0$ and $0.5$.
The achieved rates for teh decrease of the expectation
values are $\Gamma_v =0.76$\,Hz
and $\Gamma_J = 0.49$\,Hz.

\section{Conclusions and outlook}

We have extensively discussed a scheme which allows to cool simultaneously
the internal and external motion of molecules. In particular, we have
detailed the rate equations at the basis of the simulations and the
\emph{ab initio} calculations of the molecular properties. Special
care was taken to use a realistic set of parameters. For these parameters,
cooling of the external motion from 1\,K to a few $\mu\,$K can be
accomplished in about a second, thus requiring the support of trapping
technologies which are stable over these
times~\cite{Weinstein98,Drewsen04,Schiller05,BethlemNature00,RiegerPRL05,DeMilleEPJD04}.
We performed extensive calculations on the OH radical, and studied the cooling
efficiency on the molecular properties by comparing the cooling efficiency
of OH with the one of NO. We see that, even if NO has more populated
rotational levels than OH at the same initial temperature, NO can be
cooled faster because of its higher polarizability.
We notice that two properties of the polarizability play a different
role in the cooling dynamics. In particular, the absolute value of
the polarizability is responsible for the processes which cool
the rotational excitations, while for vibrational cooling
a finite slope of the polarizability as a function of the
nuclear coordinate is essential.

An interesting question is how the method performs for more complex molecules.
An advantage of larger molecules is
that states of even and odd $J$ can be coupled by Raman processes,
contrary to the case of linear molecules such as OH and NO.
On the other hand,
with increasing complexity, the heating due to spontaneous Raman scattering will
become more important and must be compensated by a faster coupling to the environment
with the help of a better cavity.
Additionally, the cooling process can be improved by optimizing the cooling sequence,
a task which becomes increasingly difficult with more complex molecules and might
require advanced optimization strategies like genetic algorithms \cite{GeneticAlgorithmExample}.
Alternatively, our technique could be extended by
combining vacuum-stimulated emission into the cavity with
suitably designed Raman pulses using optimal control
techniques~\cite{Tannor}.

\begin{acknowledgement}
G.M. acknowledges the hospitality of the
Theoretical Femtochemistry Group at LMU. Support by the European
Commission (CONQUEST, MRTN-CT-2003-505089; EMALI,
MRTN-CT-2006-035369), the Spanish MEC (Ramon-y-Cajal; Consolider
Ingenio 2010 "QOIT"; HA2005-0001),
EUROQUAM (Cavity-Mediated Molecular Cooling)
and the Deutsche
Forschungsgemeinschaft through the excellence cluster "Munich
Centre for Advanced Photonics", is acknowledged.
\end{acknowledgement}


\begin{thebibliography}{99}

\bibitem{PRL07}
G. Morigi, P.W.H. Pinkse, M. Kowalewski and R.~de~Vivie-Riedle,
Phys. Rev. Lett. accepted for publication, arXiv:quant-ph/0703157v2 (2007)

\bibitem{quovadis} see J. Doyle, B.
Friedrich, R.V. Krems, F. Masnou-Seeuws, Eur. Phys. J. D. {\bf
31}, 149 (2004), and references therein.

\bibitem{Krems05}
R.V. Krems, Int. Rev. Phys. Chem. {\bf 24}, 99 (2005)

\bibitem{Daussy99}
Ch. Daussy, T. Marrel, A. Amy-Klein, C.T. Nguyen, Ch.J. Bord\'{e},
Ch. Chardonnet, Phys. Rev. Lett. {\bf 83}, 1554 (1999)

\bibitem{Tarbutt04}
M.R. Tarbutt, H.L. Bethlem, J.J. Hudson, V.L. Ryabov, V.A. Ryzhov,
B.E. Sauer, G. Meijer, E.A. Hinds,
Phys. Rev. Lett. {\bf 92}, 173002 (2004)

\bibitem{DeMille02}
D. DeMille, Phys. Rev. Lett. {\bf 88}, 067901 (2002)

\bibitem{Tesch02}
C.M. Tesch, R.~de~Vivie-Riedle, Phys. Rev. Lett. {\bf 89}, 157901 (2002)

\bibitem{Andre06}
A. Andr\'{e}, D. DeMille, J.M. Doyle, M.D. Lukin, S.E. Maxwell, P. Rabl, R.J. Schoelkopf, P. Zoller,
Nature Phys. {\bf 9}, 636 (2006)

\bibitem{Feshbach}
D. Kleppner, Phys. Today {\bf57}(8), 12 (2004), and references therein.

\bibitem{Weinstein98} J.D. Weinstein, R. deCarvalho, T. Guillet,
B. Friedrich, J.M. Doyle, Nature (London) {\bf 395}, 148
(1998)

\bibitem{BethlemPRL99}
H.L. Bethlem, G. Berden, G. Meijer,
Phys. Rev. Lett. {\bf 83}, 1558 (1999)

\bibitem{Barker}
R. Fulton, A.I. Bishop, M.N. Shneider, P.F. Barker,
Nature Phys. {\bf 2}, 465 (2006)

\bibitem{RangwalaPRA03}
S.A. Rangwala, T. Junglen, T. Rieger, P.W.H. Pinkse, G. Rempe,
Phys. Rev. A {\bf 67}, 043406 (2003)

\bibitem{Chandler03}
M.S. Elioff, J.J. Valentini, D.W. Chandler,
Science {\bf 302}, 1940 (2003)

\bibitem{Loesch07}
Ning-Ning Liu, H.-J. Loesch
Phys. Rev. Lett. {\bf 98}, 103002 (2007)

\bibitem{Tannoudji98}
C. Cohen-Tannoudji, Rev. Mod. Phys. {\bf 70}, 707 (1998)

\bibitem{Chu98}
S. Chu, Rev. Mod. Phys. {\bf 70}, 685 (1998)

\bibitem{Phillips98}
W.D. Phillips, Rev. Mod. Phys. {\bf 70}, 721 (1998)

\bibitem{Drewsen04}
M. Drewsen, A. Mortensen, R. Martinussen, P. Staanum, J.L.
S{\o}rensen, Phys. Rev. Lett. {\bf 93}, 243201 (2004)

\bibitem{Schiller05}
P. Blythe, B. Roth, U. Fr\"ohlich, H. Wenz, S. Schiller, Phys.
Rev. Lett. {\bf 95}, 183002 (2005)

\bibitem{Stwalley} J.T. Bahns, W.C. Stwalley, P.L. Gould, J.
Chem. Phys. {\bf 104}, 9689 (1996)

\bibitem{Drewsen} I.S. Vogelius, L.B. Madsen, M. Drewsen,
Phys. Rev. A {\bf 70}, 053412 (2004)

\bibitem{Tannor}
D.J. Tannor, A. Bartana, J. Phys. Chem, \textbf{103},
10359 (1999)

\bibitem{Q-Control-Molecules} P.W. Brumer, M. Shapiro,
Principles of Quantum Control of Molecular processes, (Wiley VCH
ed., New York, 2003)

\bibitem{Horak} P. Horak, G. Hechenblaikner, K.M. Gheri, H.
Stecher, H. Ritsch, Phys. Rev. Lett. {\bf 79}, 4974 (1997)

\bibitem{Vuletic00}
V. Vuleti{\'c}, S. Chu, Phys. Rev. Lett. {\bf 84}, 3787
(2000)

\bibitem{Lev07}
B.L. Lev, A. Vukics, E. R. Hudson, B. C. Sawyer, P. Domokos, H. Ritsch, J. Ye, arXiv:0705.3639v1 (2007)

\bibitem{Lu07}
Weiping Lu, Yongkai Zhao, P.F. Barker,
Phys. Rev. A 76, 013417 (2007)

\bibitem{Maunz04} P. Maunz, T. Puppe, I. Schuster, N. Syassen,
P.W.H. Pinkse, G. Rempe, Nature (London) {\bf 428}, 50 (2004)

\bibitem{Nussmann05}
S. Nu{\ss}mann, K. Murr, M. Hijlkema, B. Weber, A. Kuhn, G. Rempe,
Nature Phys. {\bf 1}, 122 (2005)

\bibitem{Vuletic03}
H.W. Chan, A.T. Black, V. Vuletic, Phys. Rev. Let. {\bf 90} 063003 (2003)

\bibitem{DomokosRitschJOSA03} P. Domokos, H. Ritsch, J. Opt.
Soc. Am. B {\bf 20}, 1098 (2003)

\bibitem{Itano}
S. Stenholm, Rev. Mod. Phys. {\bf 58}, 699 (1986)

\bibitem{vdMeerakker05}
S.Y.T. van de Meerakker, P.H.M. Smeets, N.
Vanhaecke, R.T. Jongma, G. Meijer: Phys. Rev. Lett. {\bf 94},
023004 (2005)

\bibitem{Bochinski04} J.R. Bochinski, E.R. Hudson, H.J.
Lewandowski, J. Ye, Phys. Rev. A {\bf 70}, 043410 (2004)

\bibitem{Fulton06}
R. Fulton, A.I. Bishop, M.N. Shneider, P.F. Barker,
Nat. Phys. {\bf 2}, 465 (2006)

\bibitem{Huber77}
K.P. Huber, G. Herzberg, {\it Molecular Spectra and Molecular
Structure - IV. Constants of Diatomic Molecules}, (New York, 1977)

\bibitem{placzekteller_org}
G. Placzek, E. Teller, Z.\ f.\ Phys. {\bf 81}, 209 (1933)

\bibitem{placzekteller}
R. Gaufres, S. Sportouch, J.\ Mol.\  Spec. {\bf 39}, 527 (1971)

\bibitem{molpro}
MOLPRO, version {2006.1}, a package of {\it \emph{ab initio}} programs,
H.-J. Werner, P.J. Knowles, R. Lindh, F.R. Manby, M. Sch\"{u}tz, and
others, see http://www.molpro.net.

\bibitem{cc-pVXZ}
T.H. Dunning Jr., J. Chem. Phys. \textbf{90}, 1007 (1989)

\bibitem{linres1}
J. Olsen, P. J\o{}rgensen, J. Chem. Phys. \textbf{82},
3235 (1985)

\bibitem{linres2}
P. J\o{}rgensen, H.J.Aa. Jensen, J. Olsen,
J. Chem. Phys. \textbf{89}, 3654 (1988)

\bibitem{dalton}
DALTON, a molecular electronic structure program, Release 2.0
(2005), see http://www.kjemi.uio.no/software/dalton/dalton.html


%% Ab-intio



\bibitem{Dipmom_OH}
W. L. Meerts, A. Dymanus, Chem. Phys. Lett. \textbf{23}, 45 (1973)

\bibitem{CH2O_alpha}
J.E. Rice, R.D. Amos, S.M. Colwell, N.C. Handy, J. Sanz,
J. Chem. Phys. \textbf{93}, 8828 (1990)

\bibitem{cc-pVXZ-aug}
D.E. Woon, T.H. Dunning, J. Chem. Phys. \textbf{98}, 1358 (1993)

\bibitem{631G}
W.J. Hehre, R. Ditchfield, J.A. Pople, J. Chem. Phys. \textbf{56}, 2257 (1972)

\bibitem{Shore}
B.W. Shore, {\it The Theory of Coherent Atomic Excitation, \textbf{2}},
(Wiley-Interscience, New York, 1990)

\bibitem{Sundermann}
K. Sundermann, R. de Vivie-Riedle, J. Chem. Phys. \textbf{110},
1896 (1999)

\bibitem{BethlemNature00}
H.L. Bethlem, G. Berden, F.M.H. Crompvoets, R.T. Jongma, A.J.A.
van Roij, G. Meijer, Nature (London) {\bf 406}, 491 (2000)

\bibitem{RiegerPRL05}
T. Rieger, T. Junglen, S.A. Rangwala, P.W.H. Pinkse, G. Rempe,
Phys. Rev. Lett. {\bf 95}, 173002 (2005)


\bibitem{DeMilleEPJD04}
D. DeMille, D.R. Glenn, J. Petricka, Eur. Phys. J. D {\bf 31}, 375
(2004)

\bibitem{GeneticAlgorithmExample}
D.E. Goldberg, {\it Genetic Algorithms in Search, Optimization, and Machine Learning}
(Addison-Wesley, Reading, 1997)

\end{thebibliography}
\end{document}